\documentclass[12pt]{article}

\usepackage{fullpage}
\usepackage{epsfig}
\usepackage{latexsym}
\usepackage{verbatim}

\newcommand{\be}{\begin{equation}}
\newcommand{\ee}{\end{equation}}
\newcommand{\ba}{\begin{eqnarray}}
\newcommand{\ea}{\end{eqnarray}}
\newcommand{\bd}{\begin{displaymath}}
\newcommand{\ed}{\end{displaymath}}

\def\thalf{{\textstyle{\frac{1}{2}}}}

\def\rootg{\sqrt{-g}}

\def\DL{\mathcal{D}_L}

\begin{document}
\title{
{\bf Second order hydrodynamics for a special class of gravity duals}}

\author{{T. Springer}
\vspace*{0.1in}\\
{\it School of Physics and Astronomy}\\
{\it University of Minnesota}\\
{\it Minneapolis, Minnesota 55455, USA}}

\date{March 26, 2008}

\maketitle

\begin{abstract}
The sound mode hydrodynamic dispersion relation is
computed up to order $q^3$ for a class of gravitational duals which
includes both Schwarzschild $AdS$ and Dp-Brane metrics.  The
implications for second order transport coefficients are examined
within the context of Israel-Stewart theory.  These sound
mode results are compared with previously known results for the shear
mode.  This comparison allows one to determine the third order
hydrodynamic contributions to the shear mode for the class of metrics
considered here.
\end{abstract}

\newpage
\section{Introduction}
Hydrodynamics describes the behavior of a fluid on length and time
scales which are much longer than any microscopic scale.  The
hydrodynamic stress-energy tensor is constructed as a derivative
expansion in the fluid velocity, and is required to respect
equilibrium thermodynamics and the symmetries in the problem.  Such a
derivative expansion will contain unknown coefficients (transport
coefficients).  In first order hydrodynamics (an expansion of the
energy-momentum tensor which contains at most one derivative) without
a conserved charge, the transport coefficients which enter are the
shear viscosity $\eta$, and the bulk viscosity $\zeta$.  See
\cite{hydroreview,Kovtun2} for a more complete introduction to
hydrodynamics.

The plasma created at the Relativistic Heavy Ion Collider (RHIC)
appears to be both strongly coupled and well described by
hydrodynamics \cite{whitepapers, Molnar, Huovinen}. Transport
coefficients are necessary input for hydrodynamic simulations of the
RHIC plasma; it is desirable to calculate them, but the strong
coupling renders conventional perturbative calculations unreliable.

The AdS/CFT correspondence \cite{Maldacena, Witten, Gubser} (or more
generally, `gauge/gravity duality' or `holography') has become a
useful tool in describing strongly coupled systems, and has enjoyed
arguably its greatest success calculating hydrodynamic transport
coefficients.  There are many ways to calculate the transport coefficients of a
strongly coupled gauge theory using an extra-dimensional gravity dual.
One can compute correlation functions of the stress-energy tensor and
use Kubo formulas or examine the poles of such correlators
\cite{Policastro1, recipe,hydroI, hydroII, Herzog1, Herzog2}.
Alternatively, one can examine the behavior of the gravitational
background under perturbations and determine the dispersion relation
for such perturbations by applying appropriate boundary conditions.
Comparison with the expected dispersion relation from perturbations of
the energy-momentum tensor yields formulas for the transport
coefficients \cite{Kovtun1, Mas, ScalarSound}.  In addition, the black
hole membrane paradigm has been employed to calculate the hydrodynamic
properties of the stretched horizon of a black hole.  In many cases,
the transport coefficients calculated on the stretched horizon coincide
with the transport coefficients in the dual gauge theory
\cite{stretched, Saremi, Fujita, Kapusta, Natsuume2, Liu}.  Recently, the work
of \cite{Minwalla1} provides yet another way to compute hydrodynamic
transport coefficients by deriving the equations of fluid dynamics
directly from gravity.  This work has proved quite influential, and
has led to much subsequent research \cite{Raamsdonk, Haack, Minwalla2,
Erdmenger, Banerjee, Minwalla3}.  We use the
gravitational perturbation approach similar to \cite{Kovtun1, Mas},
as this work is an extension of a calculation in
\cite{ScalarSound}.

The quintessential example of the success of gauge/gravity duality is
the case of the shear viscosity.  In \cite{stretched} a formula for
$\eta/s$ ($s$ is the entropy density) applicable to a wide variety of
gravitational duals was derived.  Application of the formula
continually resulted in
\be
	\frac{\eta}{s} = \frac{1}{4\pi}.
\ee
It was later shown that this relation holds for all theories of
Einstein gravity \cite{Liu, BuchelLiu} (assuming the dual gauge theory
is infinitely strongly coupled).  It is quite remarkable that this
result holds for both conformal and non-conformal theories, and is
independent of the number of dimensions.  In \cite{stretched}, the
relation led to the conjecture that $\eta/s \geq 1/4\pi$ for all
substances, which has observable consequences for the plasma created
at RHIC, even though the gravitational dual to quantum chromodynamics
(QCD) is not currently known.  It is thus desirable to find other such
universal behavior from gauge/gravity duality in hopes that it may
have implications for heavy ion collisions at RHIC and at the Large
Hadron Collider (LHC).

In the past few years, much work has been done to extend previous
analyses to second order hydrodynamics \cite{Kapusta, Natsuume2,
Raamsdonk, Haack, Minwalla2, Erdmenger, Banerjee, Minwalla3,
conformalrelax1, conformalrelax2}, which attempts to repair technical
problems in first order hydrodynamics regarding causality.  Most of
the work on second order hydrodynamics so far has focused on conformal
theories.  It is notable that a universal relation between second
order hydrodynamic transport coefficients of a conformal theory was
presented in \cite{Haack2}, though it is not known whether this
relation still holds for non-conformal theories.

In this work, we examine second order hydrodynamics for a special
class of gravitational backgrounds.  This class is not necessarily
conformal, and includes both Schwarzschild AdS (conformal), and
Dp-Brane (non-conformal) backgrounds.  Specifically, we extend the
analysis of \cite{ScalarSound} to the next hydrodynamic order by
computing the sound mode dispersion relation for such a class of
backgrounds.  We also discuss implications for second order transport
coefficients within the context of the Israel-Stewart
formulation of second order hydrodynamics \cite{MIS}.

The paper is organized as follows.  In section \ref{sec1}, we detail
the specific type of gravity dual on which we focus in this work.  In
section \ref{sec2}, we add hydrodynamic perturbations on top of this
background, and review the calculation of \cite{ScalarSound}, since
this work is an extension thereof.  In section \ref{sec3}, we extend
the calculation to the next hydrodynamic order.  The hydrodynamic
dispersion relation $w(q)$ in the sound mode can be written
\be
	w(q) = w_1 q + w_2 q^2 + w_3 q^3 + ...
\ee
The main result of this section (and indeed the paper as a whole) is
the calculation of $w_3$.  In section \ref{sec4}, we discuss the
implications of our formula for $w_3$ for second order transport
coefficients
within the context of Israel-Stewart theory.  In section \ref{sec5},
we compare our results to those of a different hydrodynamic mode, the
shear mode.  It was shown in \cite{conformalrelax2} that the shear
mode dispersion relation contains contributions from (currently
unformulated) third order hydrodynamics.  Still, by comparing the
shear mode and the sound mode we can determine the value of these
extra contributions for the class of metrics we consider here.
Finally, in the last section we summarize our results and present
prospects for future investigation.

\section{Background fields}
\label{sec1}
Because this work builds upon the paper \cite{ScalarSound}, we use
this section to review the setup and notation presented therein.

\subsection{Black brane background}
In \cite{ScalarSound}, the sound mode was analyzed for black brane
gravitational backgrounds in $p+2$ dimensions.  The matter supporting
the metric was assumed to be one or more scalar fields.  The form of
the action is assumed to be
\be
	\mathcal{S} = \frac{1}{16 \pi G_{p+2}} \int \, d^{p+2}x \rootg 
	\left( R  - \thalf \sum_{k=1}^n \partial_\mu \phi_k \partial^\mu \phi_k - U(\phi_1,\phi_2...\phi_n) \right). 
\ee
The energy-momentum tensor derived from the action is
\ba
	8 \pi G_{p+2} T_{\mu \nu} &=& \thalf \sum_{k=1}^n \left(\partial_\mu \phi_k \partial_\nu \phi_k - g_{\mu \nu} \mathcal{L}_{\phi k} \right) \\
	\mathcal{L}_{\phi k} &=& \thalf \sum_{k=1}^n \partial_\lambda \phi_k \partial^\lambda \phi_k + U(\phi_1,\phi_2...\phi_n).
\ea
The metric takes the form 
\be
	ds^2 = g_{00}(r) dt^2 + g_{xx}(r) dx_j dx^j + g_{rr}(r) dr^2,
\ee
where $j = 1,2...p$, and $t$ is the time coordinate.  If this metric is presumed to be dual to a strongly coupled gauge theory
in $p+1$ dimensions, such a theory would live on the boundary at $r \rightarrow \infty$.  The position of a horizon is assumed at $r = r_0$, and
the metric components are assumed to behave in the standard way near a black brane horizon, namely that
\ba
	g_{00}(r) &\approx& -\gamma_0 (r-r_0) + \mathcal{O}(r-r_0)^2
	\label{NH}\\
	g_{rr}(r) &\approx& \frac{\gamma_r}{r-r_0} + \mathcal{O}(1) \\
	g_{xx}(r) &\approx& g_{xx}(r_0) + \mathcal{O}(r-r_0).
\ea
The quantities $\gamma_0, \gamma_r$ and $g_{xx}(r_0)$ are independent of $r$, though they
may depend on $r_0$.  The Hawking temperature of this metric is 
\be
	T = \frac{1}{4 \pi} \sqrt{\frac{\gamma_0}{\gamma_r}}.  
\ee      
For future convenience, we define the function 
\be
	F(r) \equiv -g_{00}(r)g^{xx}(r),
\ee
and we use the following notation for the logarithmic derivative
\be
	\DL [X(r)] \equiv \frac{X'(r)}{X(r)}.  
\ee
The prime denotes derivatives with respect to $r$ unless otherwise noted.  Throughout
this work, our general relativistic conventions are those of \cite{Weinberg}.

\subsection{Restrictions on the metric}
The sound mode gauge invariant equations for the type of backgrounds
mentioned above were derived in \cite{ScalarSound} in much more
generality than is necessary for this work.  Instead, we will work
with a special class of backgrounds; let us assume that the metric is
generated by a single scalar field, and that the metric components
satisfy the following constraints
\ba
	g_{rr}(r) &=& c_1 g_{xx}(r)^{p+1} F(r) \DL[F(r)]^2, \label{specialmetric1}\\
	F(r) &=& 1 - \left(\frac{g_{xx}(r_0)}{g_{xx}(r)}\right)^{c_2}.
	\label{specialmetric2}
\ea
Here $c_1, c_2$ are constants\footnote{This choice of metric is the same as the example worked out
in \cite{ScalarSound}, with the correspondence $c_2 = a_0$, and the
additional constraint $a_2 = p - a_0$ has been imposed}.  The first of
these constraints is a consequence of the fact that the particular combination
of Ricci tensor components
\be
	R^0_0 - R^x_x = 0  
\ee
for any metric that is generated by $r$ dependent scalar fields, as
shown in \cite{ScalarSound}. The second of these constraints (\ref{specialmetric2}) is
imposed to simplify the calculations.  It should be noted that 
the Schwarzschild AdS metric and the Dp-Brane metric satisfy both
constraints, and the above parametrization conveniently allows us to
compute the hydrodynamic dispersion relation for both of these
important special cases.

The Hawking temperature for this metric is given by the relation
\be
\label{specialT}
\frac{1}{(4 \pi T)^2} = c_1 g_{xx}(r_0)^p
\ee

\subsection{Scalar field and potential}
Before proceeding with the calculation, it is worthwhile to ask what
implications the above choice of metric has for the scalar field and
the scalar potential.  The combination of background Einstein equations
\be
	g^{00}G_{00} - g^{rr}G_{rr} = -8 \pi G_{p+2} \left(g^{00} T_{00} -g^{rr} T_{rr} \right) 
\ee
can be simplified to
\be
	\label{phieqn} 
	\phi'(r)^2 = 2g_{rr}\left(R^0_0(r) - R^r_r(r)\right).
\ee
Explicitly computing the right side for the special metric chosen yields a relationship between 
$\phi$ and $g_{xx}$.  
\be
	\phi(r) = \kappa \log [g_{xx}(r)] +\phi_0
\label{specialphi}
\ee
where $\phi_0$ is an integration constant, and 
\be
	\kappa \equiv \pm \sqrt{\frac{p}{2}\left(p+1-2c_2 \right)}.  
\ee

One can determine the form of the potential by considering the following combination of
background Einstein equations 
\be
	g^{00}G_{00} +  g^{rr}G_{rr} = -8 \pi G_{p+2} \left(g^{00} T_{00} + g^{rr} T_{rr} \right) = U(\phi). 
\ee
Explicitly computing the left hand side for our special metric gives
\be
	U(\phi(r)) = -\frac{p}{2 c_1\,c_2} g_{xx}(r_0)^{-2c_2} g_{xx}(r)^{2c_2-p-1} =  -\frac{p}{2 c_1\,c_2} g_{xx}(r_0)^{-2c_2} g_{xx}(r)^{-\frac{2}{p}\kappa^2}. 
\ee
Using (\ref{specialphi}), we find
\be
	U(\phi) =  -\frac{p}{2 c_1 c_2} g_{xx}(r_0)^{-2c_2} \exp \left(-\frac{2}{p}\kappa(\phi - \phi_0) \right).
\ee
Thus, the sort of metrics we are considering are those generated by a
potential which contains a single exponential, similar to the
Chamblin-Reall backgrounds \cite{Chamblin} examined in
\cite{Gubser2,Gubser3}.  Note that since the potential must be
independent of temperature, it is required that the constant $c_2$,
and the combination $c_1 g_{xx}(r_0)^{2c_2}$ must themselves be
independent of $r_0$.

\section{Hydrodynamic fluctuations}
\label{sec2}
One can access the hydrodynamic regime of the dual gauge theory by
 examining perturbations of the gravitational background, and
 following the prescription of \cite{Kovtun1}.  Here we briefly review this method
\subsection{Method for determining $w(q)$} 
In the case at hand, one should allow for fluctuations $g_{\mu \nu}
 \rightarrow g_{\mu \nu} + \delta g_{\mu \nu}$ and $\phi \rightarrow \phi +
 \delta \phi$.  The space-time dependence of the fields is presumed to
 be
\ba
	\delta g_{\mu \nu}(t,z,r) &=& e^{i(qz - wt)}h_{\mu \nu}(r), \\
	\delta \phi(t,z,r) &=& e^{i(qz - wt)} \delta \phi(r). 
\ea

Here we use the coordinate $z$ to denote one of the spatial
coordinates: $z \equiv x_p$, and $w$ and $q$ are the energy and
momentum of the perturbation.  For the sound mode, in the gauge where
$h_{\mu r} = 0$, the only non-zero fluctuations are \cite{hydroI, Mas}: 
$h_{00}(r), \frac{1}{p-1}\sum_{i=1}^{p-1} h_{ii}(r), h_{zz}(r),
h_{0z}(r)$, and $\delta \phi(r)$.

Turning on these perturbations, and expanding the background equations
of motion to first order in the perturbation leads to a set of
linearized equations for the perturbations mentioned above. To proceed, one
should take appropriate combinations of the resulting linearized
equations and construct equations involving only gauge invariant
variables, those which do not transform under the diffeomorphism
\ba
	h_{\mu \nu} &\rightarrow& h_{\mu \nu} - \nabla^{(0)}_\mu \xi_\nu - \nabla^{(0)}_\nu \xi_\mu, \\  
	(\delta \phi) &\rightarrow& (\delta \phi) - \xi_\mu (\partial^\mu \phi)
\ea
for any vector $\xi_\mu$ = $\xi_\mu(r)e^{i(qz-wt)}$.  (Here, $\nabla_{\mu}^{(0)}$ is the covariant derivative with respect
to the background metric).    

Solving the resulting gauge invariant equations perturbatively in the
hydrodynamic regime $w,q \ll T$, and imposing an incoming wave
boundary condition at the horizon and a Dirichlet boundary condition
at the boundary leads to the hydrodynamic dispersion relation $w(q)$.
One can then compare this dispersion relation to the expected
hydrodynamic form to relate the transport coefficients to the field
components in the gravity dual.

\subsection{First order hydrodynamics for this metric}
The gauge invariant equations which need to be solved for this type of metric
were derived in \cite{ScalarSound}.  They involve two gauge invariant variables
$Z_0$ and $Z_\phi$.  The equations are

\ba
	\label{z0eqn}
	\frac{g_{rr}}{\rootg} \alpha^2 F^2 \partial_r \left[\frac{\rootg g^{rr}}{\alpha^2 F^2}Z_0'\right] 
	+Z_0 \left( \DL[F]\DL[F\alpha] - g_{rr}\left(w^2g^{00} + q^2g^{xx}\right)\right) &+& \nonumber \\
	2Z_{\phi} \phi' F \left( \alpha \partial_r \left[ \frac{1}{\alpha} \left(\frac{w^2}{F} -q^2 \right) \right] 
	+ \frac{q^2 \DL[F]}{p \DL[g_{xx}]} \DL \left[ \rootg g^{rr} \phi' \right] \right)
	&=&0, \\
	\, \nonumber
\ea
\be	
	\frac{g_{rr}}{\rootg}  \partial_r \left[\rootg g^{rr}Z_\phi'\right] 
	-Z_\phi g_{rr}\left(q^2 g^{xx} + w^2 g^{00}\right) = 0,
	\label{specialZphi}
\ee
where
\be
	\alpha(r) \equiv q^2 \left((p-1)+ \frac{\DL [g_{00}(r)]}{\DL[g_{xx}(r)]}\right) - \frac{p w^2}{F(r)}.
\ee

Solving these equations perturbatively with appropriate boundary
conditions leads to the dispersion relation $w(q)$.  Classically,
waves can enter the black hole's horizon, but cannot be emitted from
there.  The standard way to apply this `incoming wave' boundary
condition is to make the ansatz \cite{Kovtun1,Mas}
\ba
	Z_0(r) &=& F(r)^{-\frac{i w(q)}{4 \pi T}} \left(Y_0(r)+ q Y_1(r)+ q^2 Y_2(r)+... \right) \\
	Z_{\phi}(r) &=& F(r)^{-\frac{i w(q)}{4 \pi T}} \left(Y_{\phi 0}(r)+ q Y_{\phi 1}(r) + q^2 Y_{\phi 2} + ... \right)\\
	w(q) &=& w_1 q + w_2 q^2 + w_3 q^3 + ...
	\label{ansatz}
\ea
The functions $Y$ must be regular at the horizon in order for the
incoming wave boundary condition to be satisfied.  One then inserts
this ansatz into the gauge invariant equations, expands the resulting
equation in powers of $q$, and solves for the functions $Y$.  Finally,
applying Dirichlet boundary conditions at the boundary ($r \rightarrow
\infty$) gives the dispersion relation.

First we solve for the functions $Y_{\phi i}$.  Inserting the above
ansatz into (\ref{specialZphi}) and expanding the result in powers of
$q$ leads to
\be
	\partial_r \left[ \rootg g^{rr} Y_{\phi 0}' \right] 
	+ q\, \partial_r \left[ \rootg g^{rr} \left( Y_{\phi 1}' - \frac{i w_1}{2 \pi T} \DL[F] Y_{\phi 0} \right) \right]+\mathcal{O}(q^2). 
\ee
It should be noted that we have omitted some terms which are
proportional to ($R^0_0 - R^x_x$) since they vanish by the background
equations of motion.  

Solving the equation for $Y_{\phi 0}$ (and using the fact that
 $\rootg g^{rr} \propto \DL[F]$ from (\ref{specialmetric1})) yields a solution
\begin{equation}
	Y_{\phi 0}(r) = k_0 + k_1 \log(F(r))
\end{equation}
but only the constant term can contribute due to the assumption of
regularity at the horizon, and presumed near horizon behavior of the
metric (\ref{NH}).  Finally the constant must be set to zero by the
Dirichlet boundary condition at infinity.  Proceeding now to higher
orders in $q$, one finds that the equations always reduce to the same
as that for $Y_0$, and as a result $Z_{\phi} = 0$ to all orders in
$q$.

Proceeding with the calculation one must now insert the incoming wave
ansatz into the remaining gauge invariant equation (\ref{z0eqn}),
\be	
	\label{specialz0eqn}
	\frac{g_{rr}}{\rootg} \alpha^2 F^2 \partial_r \left[\frac{\rootg g^{rr}}{\alpha^2 F^2}Z_0'\right] 
	+Z_0 \left[ \DL[F]\DL[F\alpha] - g_{rr}\left(w^2g^{00} + q^2g^{xx}\right)\right]
	=0,
\ee
solve for the $Y_i$ functions, and impose regularity at the horizon and
a Dirichlet boundary condition at $r \rightarrow \infty$.  These steps
were completed up to $\mathcal{O}(q)$ in \cite{ScalarSound}.  The
results for the type of metric we consider here are summarized below.
\ba
	Y_0(r) &=& y_0\left(F(r)-1 \right),
	\label{y0soln}\\
	Y_1(r) &=& y_1\left(F(r)-1 \right),
	\label{y1soln}\\
	w_1 &=& \pm \sqrt{\frac{2c_2}{p}-1} ,
	\label{w1soln}\\
	w_2 &=& \frac{-i(p-c_2)}{2 \pi T p}.
	\label{w2soln}
\ea
Here $y_0$ and $y_1$ are constants. 

Comparing these results to the dispersion relation expected from first order hydrodynamics
\ba
	w_1 &=& \pm v_s, \\
	w_2 &=& - i \frac{\eta}{\epsilon + P} \left( \frac{p-1}{p} + \frac{\zeta}{2\eta} \right),
\ea
one gains knowledge of $v_s$ (speed of sound), $\eta$ (shear viscosity), and $\zeta$ (bulk viscosity).
\ba
	\eta/s &=& 1/4\pi \label{etasoln}\\
	v_s &=& \sqrt{\frac{2c_2}{p}-1} \label{vssoln}\\
	\zeta / \eta &=& 2 \left(\frac{1}{p} - v_s^2 \right) \label{zetasoln}.
\ea
Here $s = (\epsilon + P)/T$ is the entropy density,  $\epsilon$ is the
equilibrium energy density, and $P$ is the equilibrium pressure.  Note
that the conjectured bulk viscosity bound of Buchel \cite{Buchel} is
saturated for metrics of the type we consider.  It was also shown in
\cite{ScalarSound} that the above results agree with previous
calculations for the Schwarzschild AdS metric (with the choice $c_2 =
(p+1)/2$.  These results also agree with the calculation of \cite{Mas}
for the Dp-Brane metric with the choice $c_2 = p(7-p)/(9-p)$.
	  
\section{Solution for $w_3$}
\label{sec3}
It is the purpose of this work to extend the above calculation to the next hydrodynamic order, 
and thus to determine the next coefficient in the dispersion relation $w_3$. 
\subsection{Equation for $Y_2$}

We have already shown that $Z_{\phi}$ vanishes to all orders in $q$,
so then it remains to return to (\ref{specialz0eqn}), insert the
incoming wave ansatz, expand in powers of $q$, and insert the
solutions (\ref{y0soln} - \ref{w2soln}).  \ Completing these steps,
one finds the following differential equation which must be solved for
$Y_2$:
\ba
	&\,& \partial_r \left[ \frac{ Y_2'}{\DL[F](1+F)^2} \right] + \nonumber \\ 
	&\,& \frac{F'}{(1+F)^3}\left\{Y_2 +\frac{y_0}{(4 \pi T)^2} \left[x_0 
	+ \frac{1-F^2}{F} \left(w_1^2 + \frac{g_{xx}(r)^p}{g_{xx}(r_0)^p} \left(F-w_1^2 \right) \right)\right]\right\} = 0.
\label{y2eqn}
\ea
where
\be
	x_0 \equiv 2 \left( 5w_1^2-1 - \frac{2 w_1 w_3 (4 \pi T)^2}{1-w_1^2} \right).
	\label{x0definition}
\ee
In writing the above expression, we have replaced all occurrences of
the constant $c_2$ which appears in the metric with $w_1$ due to the
relation (\ref{w1soln}), and have removed the constant $c_1$ in favor of $T$ using (\ref{specialT}).  

\subsection{Solution for $Y_2$}  
\subsubsection{Solution to the associated homogeneous equation}
The associated homogeneous equation for (\ref{y2eqn}) is the same as
the homogeneous part of the equation for $Y_1$ which was considered in
\cite{ScalarSound}.  The solution can easily be found by first
making the ansatz $Y_2(r) = y_{2a} (1-F(r))$, and then using the technique of
reduction of order once this solution is found.  The general
solution to the homogeneous part of the above equation is
\be
	Y_{2h}(r) = y_{2a} (1 - F(r)) + y_{2b} \Bigl[ (1-F(r))\log[F(r)] + 4 \Bigr].
\ee
Here the subscript $h$ stands for homogeneous, and $y_{2a}$, $y_{2b}$ are arbitrary constants.  
\subsubsection{Particular solution}
In order to find the general solution for $Y_2$, we must now find a
particular solution to the inhomogeneous equation.  Since the
homogeneous solution is known, one can construct a particular solution
using the method of variation of parameters.  For completeness, this
method is outlined in Appendix \ref{VarOfParams}. Applying the method
to the case at hand gives the following solution (details are
given in Appendix \ref{y2pAppendix}).  
\ba
	Y_{2p}(r) &=& - \frac{x_0 y_0}{(4 \pi T)^2} + \frac{y_0}{(4 \pi T)^2} ( F(r) - 1 ) \left[ \int \left(\frac{1+F(r)}{1-F(r)}\right)^2 \frac{F'(r)}{F(r)} \hat{Q}(r) \, dr \right] \\
	\hat{Q}(r) &=& \int^r \left(\frac{1-F(\tilde{r})}{1+F(\tilde{r})}\right)^2 \frac{F'(\tilde{r})}{F(\tilde{r})}\left[w_1^2 + \frac{g_{xx}(\tilde{r})^p}{g_{xx}(r_0)^p} \left(F(\tilde{r})-w_1^2 \right)\right] \,d\tilde{r},
\label {yp}
\ea
Thus, the general solution for $Y_2$ is given by
\be
	Y_2(r) = Y_{2h}(r) + Y_{2p}(r).  
\ee
To proceed, it is convenient to change coordinates.  We define the coordinate $u$ by
\be
	u^2 \equiv \left(\frac{g_{xx}(r_0)}{g_{xx}(r)}\right)^{c_2} = \left(\frac{g_{xx}(r_0)}{g_{xx}(r)}\right)^{\frac{p}{2}(w_1^2+1)} ,
\ee
so that
\be
	F(u) = 1 - u^2,
\ee
and the horizon is now located at $u = 1$, and the boundary is located
at $u=0$.\footnote{Here the assumption is that $c_2 > 0$, and that
$g_{xx}(r\rightarrow \infty) \sim
r^n$ with $n>0$  These assumptions hold for the Schwarzschild $AdS$
metric for any positive $p$, and for the Dp-Brane metric provided
$p<7$.}  In terms of the new coordinates, $Y_2$ is given by
\ba
	Y_{2h}(u) &=& y_{2a} u^2 + y_{2b} \Bigl[ u^2\log[1-u^2] + 4 \Bigr],\\
	Y_{2p}(u) &=& -\frac{y_0}{(4 \pi T)^2} \left \{x_0 + 4 u^2 \int \frac{(2-u^2)^2}{u^3(1-u^2)} 
		\int^u \frac{v^5\left[w_1^2+v^{-4/(1+w_1^2)}\left(1-v^2-w_1^2\right) \right]}{(2-v^2)^2(1-v^2)}\,dv\,du \right\} \nonumber \\
		\,
\ea
The inner integral can be written in terms of Hypergeometric
functions, but the result is not particularly enlightening, so we do
not reproduce it here.  The full analytical form of this function is
not needed to determine the dispersion relation.

\subsection{Boundary conditions}
\subsubsection{Regularity at horizon}
The first boundary condition which must be applied on $Y_2$ is the
condition of regularity at the horizon.  In order to do so, one must
extract the coefficient of the logarithmic divergence in the
particular solution $Y_{2p}$.  To do so, we first expand the integrand
in powers of $(u-1)$, and look for the coefficient of the $(u-1)^{-1}$
term.  After integration, this term will lead to the logarithmic
divergence.

With the aid of Mathematica, we find the nested integral can be expanded near
the horizon as
\ba
&\,&\int^u \frac{v^5\left[w_1^2+v^{-4/(1+w_1^2)}\left(1-v^2-w_1^2\right) \right]}{(2-v^2)^2(1-v^2)}\,dv \nonumber \\
&\approx& \frac{1}{2}\left\{\frac{1-3w_1^2}{1-w_1^2} - w_1^2\left[2 + i \pi - H_n\left(\frac{-2}{1+w_1^2}\right) \right]\right\} + \mathcal{O}(u-1)
\ea
where $H_n(\alpha)$ is the `Harmonic Number' defined as
\be
	H_n(\alpha) \equiv \int_0^1 \frac{1-x^{\alpha}}{1-x} \,dx.
\ee
Using this expansion in $Y_2$, we find that near the horizon, 
\ba
	&\,&Y_{2p}(u \rightarrow 1) \\
	&\approx& -\frac{y_0}{(4 \pi T)^2} \left\{x_0 + 4 \int 
	\frac{du}{4(1-u)}\left[\frac{1-3w_1^2}{1-w_1^2} - w_1^2\left(2 + i \pi - H_n\left(\frac{-2}{1+w_1^2}\right) \right)\right] + \mathcal{O}(1) \right\}. \nonumber
\ea
Going back to the general solution for $Y_2$, one thus finds
\ba
	&\,& Y_{2}(u \rightarrow 1) \\
	&\approx& \log [1-u] \left\{ y_{2b} +\frac{y_0}{(4 \pi T)^2}   
	\left[\frac{1-3w_1^2}{1-w_1^2} - w_1^2\left(2 + i \pi - H_n\left(\frac{-2}{1+w_1^2}\right) \right)\right] \right\} + \mathcal{O}(1). \nonumber
\ea
The requirement of regularity at $u = 1$ thus gives
\be
        y_{2b} = \frac{y_0}{(4 \pi T)^2}   
	\left[ w_1^2\left(2 + i \pi - H_n\left(\frac{-2}{1+w_1^2}\right) \right)- \frac{1-3w_1^2}{1-w_1^2} \right].
\ee
For future convenience, we make use of the identity
\be
	H_n(\alpha)=H_n(\alpha+1) - \frac{1}{1+\alpha} 
\label{Hnidentity}
\ee
to write
\be
	H_n\left(\frac{-2}{1+w_1^2}\right) = H_n\left(\frac{2w_1^2}{1+w_1^2}\right) - \frac{1-2w_1^2-3w_1^4}{2w_1^2(1-w_1^2)}.
\ee
Substituting this into the equation for $y_{2b}$ we find
\be
	y_{2b} =  \frac{y_0}{(4 \pi T)^2} 
	\left[w_1^2\left(2 + i\pi -H_n\left(\frac{2w_1^2}{1+w_1^2}\right)\right) -\frac{1}{2} (1-3 w_1^2) \right].  
	\label{y2bsoln}
\ee

\subsubsection{Dirichlet boundary condition at $u = 0$}
Finally, we must apply the Dirichlet boundary condition at $u = 0$.  We proceed as above, by expanding the integrand of $Y_{2p}$ near 
$u = 0$.  Mathematica gives
\be
\int^u \frac{v^5\left[w_1^2+v^{-4/(1+w_1^2)}\left(1-v^2-w_1^2\right) \right]}{(2-v^2)^2(1-v^2)}\,dv \approx -\frac{w_1^2}{2}\left(2+i \pi\right) + \mathcal{O}(u),
\ee
so that 
\be
	Y_{2}(u \rightarrow 0) \approx \Bigl( 4 y_{2b} + \mathcal{O}(u)\Bigr) - \frac{y_0}{(4 \pi T)^2} \left[ x_0 - 8 u^2 w_1^2(2+i\pi) \int \left[\frac{1}{u^3} + \mathcal{O}(u^{-2})\right]du \right].
\ee
Doing the integral, one finally has
\be
	 Y_{2}(u \rightarrow 0) \approx 4 y_{2b} - \frac{y_0}{(4 \pi T)^2} \left[ x_0 + 4 w_1^2 \left(2+i \pi\right)\right] + \mathcal{O}(u),
\ee
and applying $Y_2(u \rightarrow 0) = 0$ gives
\be
	4 y_{2b} = \frac{y_0}{(4 \pi T)^2} \left[ x_0 + 4 w_1^2 \left(2+i \pi\right) \right],
\ee
Using (\ref{y2bsoln}),
\be
	x_0 = -4 \left[w_1^2 H_n\left(\frac{2w_1^2}{1+w_1^2}\right) + \frac{1}{2}(1-3w_1^2)\right].
\ee
Finally, (\ref{x0definition}) allows us to solve for $w_3$.  The result is
\be
	w_3 = \frac{w_1(1-w_1^2)}{(4 \pi T)^2} \left[1+  H_n\left(\frac{2w_1^2}{1+w_1^2}\right) \right]. 
	\label{w3soln}
\ee
This is our main result.  To summarize, we have computed the
coefficient of the $q^3$ term in the sound mode hydrodynamic
dispersion relation for a specific class of metrics (see
(\ref{specialmetric2})).  This class of metrics contains two constants
which we denote $c_1$ and $c_2$ .  Our expression for $w_3$ should
necessarily depend on these constants, but we have eliminated $c_1$ in
favor of $T$ using (\ref{specialT}) and $c_2$ in favor of $w_1^2 =
v_s^2$ due to the relation (\ref{vssoln}).

\section{Transport coefficients in Israel-Stewart theory}
\label{sec4}
Now that we have computed the dispersion relation to
$\mathcal{O}(q^3)$ in the sound mode, we are in a position to examine
the implications for second order hydrodynamic transport coefficients.
Second order hydrodynamics attempts to repair some technical problems
regarding causality in the first order theory.  This subject was first
broached by M\"{u}ller \cite{Muller}, and later by Israel and Stewart
\cite{MIS}.  Recently, the formulation of second order hydrodynamics
presented in \cite{conformalrelax2} has gained popularity, though at
present it is only applicable to conformal theories.  (It should be
noted that recently, some progress has been made in generalizing the
work of \cite{conformalrelax2} to non-conformal theories
\cite{Skenderis}).  The metrics we consider are not necessarily
conformal, and thus we will use the Israel-Stewart formulation.

Israel introduced five new transport coefficients that appear
in the hydrodynamic expansion of the energy momentum tensor.  In what
follows, we use the same notations and conventions as
\cite{conformalrelax1}.  Three of these five transport coefficients
are relaxation times associated with the diffusive, shear, and sound
mode, and are denoted by $(\tau_J, \tau_{\pi}, \tau_{\Pi})$
respectively.  There are two other transport coefficients which are
related to coupling between the different modes, $\alpha_0, \alpha_1$.

In \cite{conformalrelax1}, the sound mode dispersion relation was computed in terms of these
transport coefficients. 

\ba
	w_1 &=& \pm v_s \\
	w_2 &=& \frac{i}{T s}\left(\frac{p-1}{p} \eta + \frac{\zeta}{2}\right)\\
	w_3 &=& \pm \frac{\eta }{2 v_s T s}\left[\frac{p-1}{p}\left(2 v_s^2 \tau_{\pi} - \left(1-\frac{1}{p}\right)\frac{\eta }{T s}\right)+\frac{\zeta }{\eta }\left(v_s^2 \tau_{\Pi}- \left(1-\frac{1}{p}\right)\frac{\eta }{T s}-\frac{\zeta }{4 T s}\right)\right]
	\label{w3transport}
\ea
These relations were derived within the context of a `decoupled
ansatz' which presumes the background contains no R-charge.  The cases
which we will consider below (Schwarzschild AdS and Dp-Brane) fit this
criteria.  Our backgrounds are generated by scalar fields only, and
thus any gauge field necessary to provide R-charge is absent from the
cases we consider here.   A more complete list of
assumptions regarding this dispersion relation can be found in
\cite{conformalrelax1}.

Comparing (\ref{w3transport}) to our main result (\ref{w3soln}), and
eliminating $\eta$ and $\zeta$ from the relations
(\ref{etasoln},\ref{zetasoln}) gives the relation
\be
	\tau_{\pi}+ \frac{\left(1-p v_s^2 \right)}{(p-1)} \tau_{\Pi} - \frac{\left(1-v_s^2\right)}{(8 \pi  T)v_s^2}\frac{p}{(p-1)}\left\{1+v_s^2\left[1+ 2 H_n \left(\frac{2 v_s^2}{1+v_s^2} \right)\right]\right\} = 0	
\label{taupirelation}
\ee
As expected, the coefficients $\tau_{\pi}$ and $\tau_{\Pi}$ cannot in
general be determined separately using this method.  Still, if one of
these coefficients is known, the above relation allows us to determine
the other.

Let us now explicitly check that our results agree with other
calculations in the case of a conformal background.  The Schwarzschild
AdS black hole metric in the near horizon limit takes the form
\ba
	ds^2 &=& \frac{r^2}{L^2} \left[ -F(r) dt^2 + dx_j dx^j \right] + \frac{L^2 dr^2}{r^2 F(r)}\\
	F(r) &=& 1 - \left(\frac{r_0}{r} \right)^{p+1}
\ea
where $L$ is the radius of curvature of the $AdS$ space.  For this
metric, the parameter $c_2 = (p+1)/2$ and thus $v_s^2 = 1/p$.  In this
case, (\ref{taupirelation}) gives
\be
	\tau_{\pi}^{SAdS} = \frac{1}{4 \pi T} \left[ \frac{p+1}{2} + H_n\left(\frac{2}{p+1} \right) \right],  	
\label{taupiconformal}
\ee  
which is in agreement with \cite{Haack}, \cite{Minwalla2}\footnote{In
comparing with the results of \cite{Minwalla2}, one needs to employ
the identity (\ref{Hnidentity}) to see the agreement}.  This is a
non-trivial check on our calculation; the cited results were arrived
at by completely different methods than those we employ here.
Furthermore, it should be noted that (\ref{taupiconformal}) confirms a
conjecture made by Natsuume in \cite{Natsuume2}\footnote{To be precise, the
conjecture is confirmed for the case of $p \geq 2$; the case of $p=1$
should probably be checked separately as the derivation of the gauge
invariant equations in \cite{Mas}, \cite{ScalarSound} rely on at least
2 spatial dimensions.  See \cite{D1brane}, where first order
hydrodynamics is examined for $p=1$}.

Finally, we can also consider the case of the Dp-Branes.  In the Einstein frame, the metric can be 
reduced to \cite{stretched, Mas}
\ba
	ds^2 &=& \left(\frac{r}{L}\right)^\frac{9-p}{p}\left[-F(r) dt^2+dx_jdx^j\right] 
	+ \left( \frac{r}{L} \right)^{\frac{p^2 - 8p +9}{p}}\frac{dr^2}{F(r)}, \\
	F(r) &=& 1-\left(\frac{r_0}{r}\right)^{7-p}.  
\ea
for this metric, the parameter $c_2 = (7p-p^2)/(9-p)$, and $v_s^2 = (5-p)/(9-p)$ \cite{Mas}.  Inserting
this into (\ref{taupirelation}) one finds
\be
	(9-p)(1-p) \tau_{\pi}^{Dp} = (3-p)^2 \tau_{\Pi}^{Dp} -\frac{p}{ \pi  T}\frac{(7-p)}{(5-p)}\left(1+\frac{5-p}{7-p} H_n\left(\frac{5-p}{7-p}\right)\right).  
\ee
This formula agrees with previous computations for $p=1$ and $p=4$ in \cite{conformalrelax1}.

\section{Comparison with the shear mode}
\label{sec5}
Though we have exclusively dealt with the sound mode in this work, we
now compare our results to similar ones from the shear mode.  Let us
define the shear mode dispersion relation in the same way as
\cite{conformalrelax1},
\be
	w(q)_{\rm shear} = -i D_{\eta} q^2 - i D_{\eta}^2 \tau_{\rm shear} q^4 + \mathcal{O}(q^6)  
\ee
where 
\be
	D_{\eta} = \frac{\eta}{T s} = \frac{1}{4 \pi T}.
\ee
A formula for $\tau_{\rm shear}$ was given in \cite{Kapusta} which is
applicable to a wide variety of metrics, including the special metrics
we have considered in this note.  It states that
\be
	\tau_{\rm shear} = \frac{\sqrt{-g(r_0)}}{\sqrt{-g_{00}(r_0) g_{rr}(r_0)}}
	\int_{r_0}^\infty dr \frac{g_{rr}(r)}{\sqrt{-g(r)}}
	\left[ 1- \left( \frac{D(r)}{D(r_0)} \right)^2 \right]
	\label{relaxation}
\ee
where
\be
	D(r) \equiv \frac{\sqrt{-g(r)}}{\sqrt{-g_{00}(r)g_{rr}(r)}}
	\int_r^\infty dr^\prime 
	\frac{-g_{00}(r^\prime)g_{rr}(r^\prime)}
	{\sqrt{-g(r^\prime)}g_{xx}(r^\prime)} \, .
\ee
Using the special metric (\ref{specialmetric1}-\ref{specialmetric2}),
and the relationship (\ref{specialT}) in this formula yields
\be
	\tau_{\rm shear} = \frac{1}{4 \pi T} H_n \left(2 - \frac{p}{c_2}\right) = \frac{1}{4 \pi T} H_n \left(\frac{2w_1^2}{1+w_1^2}\right).
\label{taushear}
\ee 
This result agrees
with the special cases of the Dp-Brane and the Schwarzschild AdS
metrics as computed in \cite{Natsuume2}.

Previously, it was thought that $\tau_{\rm shear} = \tau_{\pi}$, but
recently the authors of \cite{conformalrelax2} showed that this is not
the case because the coefficient of the $q^4$ term contains not only
$\tau_\pi$, but also contributions from (currently unformulated) third
order hydrodynamics.  We can now determine these unknown contributions
for metrics which obey (\ref{specialmetric1},\ref{specialmetric2}).

Let us parametrize 
\be
	\tau_{\rm shear} = \tau_{\pi} + \Delta
\ee
where $\Delta$ denotes the unknown contributions from third order
hydrodynamics.  Combining (\ref{taushear}) and (\ref{taupirelation})
allows one to solve for $\Delta$.  Evidently,
\be
	\Delta = \left(\frac{p v_s^2-1}{p\left(1-v_s^2\right)}\right)(\tau_{\pi}-\tau_{\Pi}) -\frac{1+v_s^2}{8 \pi  T v_s^2}. 
\label{deltasoln}
\ee
The correction $\Delta$ does not appear to be universal in the sense
that the first term is not present in the case of a conformal theory.
In the conformal case, $\Delta$ is still not `universal', because it
depends on the number of dimensions of the theory.  In the future,
when the particular transport coefficients which comprise $\Delta$ are
known, it will be interesting to see whether there is any universal
relationship between these unknown coefficients, $\tau_{\pi}$, and
$\tau_{\Pi}$.

We can easily check that the formula (\ref{deltasoln}) reproduces the
results in the well known $SAdS_5$ metric, which is dual to
$\mathcal{N}=4$ supersymmetric Yang-Mills theory at finite
temperature.  In this case, $p=3$ and $v_s^2 = 1/3$.  Immediately, we
have
\be
	\Delta^{SADS5} = - \frac{1}{2 \pi T}.  
\ee
In \cite{conformalrelax1,conformalrelax2}, it was found
that\footnote{Even though the formalism of Baier et
al. \cite{conformalrelax1} is different than Israel-Stewart, one can
check that the sound mode dispersion relations coincide in the
limit of conformal theories ($\zeta \rightarrow 0$).  In an
unfortunate clash of notations, the relaxation time introduced by Baier
et al. is denoted by $\tau_\Pi$, its Israel-Stewart counterpart is
$\tau_\pi$.}
\ba
	\tau^{SADS5}_{\rm shear} &=& \frac{1-\log (2) }{2 \pi T} \\
	\tau^{SADS5}_{\pi} &=& \frac{2- \log(2) }{2 \pi T}.
\ea
It is clear that the relaxation time computed from the shear mode and the sound
mode differ by the amount predicted by the formula (\ref{deltasoln}); our results are in agreement with \cite{conformalrelax1, conformalrelax2}.  

\section{Conclusion and outlook}
In this work, we have extended the calculation of \cite{ScalarSound}
to the next hydrodynamic order in $q$.  The main result (\ref{w3soln})
is the coefficient of the $q^3$ term in the sound mode dispersion
relation.  This result is applicable to metrics which obey
(\ref{specialmetric1} - \ref{specialmetric2}).  These metrics are not
necessarily conformal, and contain an arbitrary number of spatial
dimensions $p$ with $p>2$.

Information about second order transport coefficients was presented
within the context of the Israel-Stewart theory (specifically within
the formulation presented in \cite{conformalrelax1}).  In general, a
relationship (\ref{taupirelation}) between two transport coefficients
$\tau_{\pi}$ and $\tau_{\Pi}$ can be determined.

In the conformal case of the Schwarzschild $AdS$ metric, the relation
mentioned above allows the determination of the coefficient
$\tau_{\pi}$.  We have verified that our results agree with those
calculated from different methods.

Finally, by comparing the sound mode dispersion relation to the shear
mode discussed in \cite{Kapusta, Natsuume2}, we were able to determine
the contribution of third order hydrodynamics to the shear mode
(it was pointed out that such contributions would be present in
\cite{conformalrelax2}).

As mentioned in the introduction, it is desirable to find other
universal relations among transport coefficients, as such relations
sometimes lead to observable consequences.  The main results of this
paper (\ref{w3soln}), (\ref{taupirelation}) and the third order
hydrodynamic contributions to the shear mode (\ref{deltasoln}) are
applicable to certain gravity duals which may or may not be conformal.  
However, despite the fact that
these relations appear to be applicable to many theories, they do not
seem to be universal in the same way as $\eta/s = 1/4\pi$, and the
relation presented in \cite{Haack2}.  For example, the number of
spatial dimensions $p$ enters explicitly into our formulas, whereas a
universal relation should not depend on this quantity.

Furthermore, we stress that the class of theories examined herein is
quite limited.  It seems likely that any generalization will
explicitly contain the bulk viscosity $\zeta$, which we were able to
eliminate due to the relation (\ref{zetasoln}).

Of course, it would be interesting to re-examine the results here for
a broader class of gravity duals.  Unfortunately, the gauge invariant
equations derived in \cite{ScalarSound} are difficult to solve
analytically except in the special case presented here, though perhaps
insight could be gained by approaching the problem numerically.  It
would also be useful to attempt to generalize the results of
\cite{ScalarSound} by including other matter fields in addition to
the scalar fields considered there.  It seems unlikely (though perhaps
not impossible) that similar special cases would be analytically
solvable after the addition of different kinds of matter.  These are issues
which should be investigated in the future.  

\section{Acknowledgments}
I would like to thank Joe Kapusta, Alex Buchel, Makoto Natsuume,
Aleksey Cherman, and Mitsutoshi Fujita for helpful comments and
discussions.  This work was supported by the US Department of Energy
(DOE) under Grant. No. DE-FG02-87ER40328, and by the Graduate School
at the University of Minnesota under the Doctoral Dissertation
Fellowship.

\appendix
\section*{Appendices}
\section{The method of variation of parameters}
\label{VarOfParams}

One fundamental technique used in generating particular solutions to
inhomogeneous differential equations is the method of `Variation of
Parameters'.  The method can be used to find a general solution to a
second order linear differential equation if the solution to the
associated homogeneous equation is known.

The theory behind the method can be found in any textbook on
differential equations.  Here, we simply present the essential formulas.
Consider a differential equation
\be
	y''(x) + p(x) y'(x) + q(x) y(x) = g(x),
\ee
and assume that the functions $y_1(x)$ and $y_2(x)$ are linearly
independent, and satisfy the associated homogeneous equation.  That
is,
\be
	y_1''(x) + p(x)y_1'(x) + q(x)y_1'(x) = 0,
\ee
and similarly for $y_2(x)$.  It can be shown that the function
$y_p(x)$ is a solution to the inhomogeneous equation, where
\be
	y_p(x) = y_2(x) \int \frac{y_1(x)g(x)}{W(x)} \,dx -  y_1(x) \int \frac{y_2(x)g(x)}{W(x)}.
\ee
Here, $W(x)$ is the Wronskian
\be
	W(x) \equiv y_1(x)y_2'(x) - y_1'(x)y_2(x).
\ee
To arrive at the form of the function $y_p$ used in the text (\ref{yp}), let us define
\be
	h(x) \equiv \frac{y_2(x)}{y_1(x)}.  
\ee 
Then,
\be
	W(x) =  h'(x)y_1^2(x),  
\ee
and
\be
	y_p(x) = y_1(x)\left[h(x) \int \frac{g(x)}{y_1(x)h'(x)} \,dx - \int \frac{h(x)g(x)}{h'(x)y_1(x)}\right].
\ee
One can see that this is equivalent to
\be
	y_p(x) = y_1(x) \int h'(x) \int^x \frac{g(z)}{y_1(z)h'(z)} \,dz \,dx
\ee
by performing an integration by parts. 

\section{Particular solution for $Y_2(r)$}
\label{y2pAppendix}
One can now apply the method detailed in Appendix \ref{VarOfParams} to
the differential equation for $Y_2$.  Using the notation from Appendix
\ref{VarOfParams}, (and changing independent variables from $x$ to
$r$), we have
\ba
	y_1(r) &=& 1- F(r),\\
	h(r) &=& \log[F(r)] + \frac{4}{1-F(r)},\\
	g(r) &=& -\frac{(F')^2}{F(1+F)} \frac{y_0}{(4 \pi T)^2} \left[x_0 
	+ \frac{1-F^2}{F} \left(w_1^2 + \frac{g_{xx}(r)^p}{g_{xx}(r_0)^p} \left(F-w_1^2 \right) \right)\right].
\ea
After a bit of work, one can find the particular solution
\ba
	Y_{2p}(r) &=& -\frac{y_0}{(4 \pi T)^2} (1-F(r)) \left[ \int \left(\frac{1+F(r)}{1-F(r)}\right)^2 \frac{F'(r)}{F(r)} Q(r) \, dr \right] \\
	Q(r) &=& \int^r \frac{(1-F(z))}{(1+F(z))^3} F'(z)\left[x_0 
	+ \frac{1-F(z)^2}{F(z)} \left(w_1^2 + \frac{g_{xx}(z)^p}{g_{xx}(r_0)^p} \left(F(z)-w_1^2 \right) \right)\right] \,dz. \nonumber \\
\,   
\ea
The term involving $x_0$ can be integrated directly, by changing
variables from $z$ to $F$.  This simplifies the solution to
\ba
	Y_{2p}(r) &=& - \frac{x_0 y_0}{(4 \pi T)^2} + \frac{y_0}{(4 \pi T)^2} ( F(r) - 1 ) \left[ \int \left(\frac{1+F(r)}{1-F(r)}\right)^2 \frac{F'(r)}{F(r)} \hat{Q}(r) \, dr \right] \\
	\hat{Q}(r) &=& \int^r \left(\frac{1-F(z)}{1+F(z)}\right)^2 \frac{F'(z)}{F(z)}\left[w_1^2 + \frac{g_{xx}(z)^p}{g_{xx}(r_0)^p} \left(F(z)-w_1^2 \right)\right] \,dz,
\ea
which is the form presented in the text.  

\end{document}